\documentclass[useAMS,usenatbib,final]{mn2e}

\usepackage{amsmath}

\usepackage{ifpdf}

\ifpdf
\usepackage[pdftex]{graphicx}
\else
\usepackage{graphicx}
\fi

\ifpdf
\DeclareGraphicsExtensions{.pdf,.png,.jpg}
\else
\DeclareGraphicsExtensions{.eps,.ps}
\fi

\newcommand{\hd}{HD\,149026\,b}
\newcommand{\hdA}{HD\,149026}

\title[The formation of \hd]{The formation of \hd}
\author[C. Broeg and G. Wuchterl]{C. Broeg$^{1,2,3}$\thanks{E-mail:
broeg@space.unibe.ch} and G. Wuchterl$^{3}$\\
$^{1}$ Astrophysical Institute and Obervatory, Friedrich-Schiller
University, Schillerg\"a\ss chen 2-3, Jena, 07745, Germany\\
$^{2}$ Max-Planck Institute for extraterrestrial Physics,
Giessenbachstrasse, Garching, 85741, Germany\\
$^{3}$ Th\"uringer Landessternwarte,  Sternwarte 5,
Tautenburg, 07778, Germany}

\begin{document}
\date{Accepted 2007 Jan 9. Received 2006 Dec 18; in original form 2006
  September 27}

\pagerange{\pageref{firstpage}--\pageref{lastpage}} \pubyear{2002}

\maketitle

\label{firstpage}

\begin{abstract}
Today, many extrasolar planets have been detected. Some of them
exhibit properties quite different from the planets in our solar
system and they have eluded attempts to explain
their formation. One such case is \hd. It was discovered by
\citet{2005ApJ...633..465S}. A transit-determined
orbital inclination results in a total mass of $114\, \mathrm{M_{\earth}}$.
The unusually small radius can be explained by a condensible
element core with an inferred mass of $67\,\mathrm{M_{\earth}}$ for the best fitting
theoretical model.

In the core accretion model, giant planets are assumed to form around
a growing core of condensible materials.
With increasing core mass, the amount of gravitationally bound envelope mass increases. This continues up to the so-called critical core mass -- the largest core allowing a hydrostatic envelope. For larger cores, the lack of static solutions  forces a dynamic evolution of the protoplanet in the process accreting large amounts of gas or ejecting the envelope. This would prevent the formation of \hd.

By studying all possible hydrostatic equilibria we could show that \hd\ can
remain hydrostatic up to the inferred heavy core. This is possible if it is formed in-situ in a
relatively low-pressure nebula. This formation process is
confirmed by fluid-dynamic calculations using the environmental
conditions as determined by the hydrostatic models.

We present a quantitative in-situ formation scenario for the
massive core planet \hd. Furthermore we predict a wide range of
possible core masses for close-in planets like \hd. This is different
from migration where typical critical core masses should be expected.
\end{abstract}


\section{Introduction}
At the moment, there are two competing theories for giant planet
formation. In one theory, the
solar nebula fragments directly due to gravitational instability to
form a giant planet \citep{1978M&P....18....5C,1979Icar...38..367D}. 
In the second, the core accretion scenario, a core or protoplanet
forms through accretion of planetesimals and when its
mass reaches some critical value the surrounding gaseous envelope
is supposed to become unstable and to collapse onto the core, in the process forming
the giant planet \citep{1974Icar...22..416P,1978PThPh..60..699M}. For
an in-depth discussion see \citet{2000prpl.conf.1081W}.
In
this paper, we will assume that the nebula is gravitationally stable 
and follow the second idea.

Soon after the discovery of the first exo-planet, 51~Peg\,b, by
\citet{1995Natur.378..355M} 
theorists have come up with several migration theories predicting that the
planets form at large orbital distances (around 4-5 AU) and migrate
inwards. Both continuous \citep{1996Natur.380..606L} and sudden
migration \citep[the jumping Jupiters of][]{1977Ap&SS..51..153W} have
been proposed.
On the other hand, \citet{1996DPS....28.1107W,1997svlt.work...64W}
	    showed that in-situ formation could occur if sufficent amounts
	    of solids and gas are available in the planets feeding-zone.

The planet \hd\ was discovered by \citet{2005ApJ...633..465S} at a
distance $a$ of only 0.042\,AU. Because it was discovered by both the
radial velocity and the transit method, its mass and density are
known. It has a total mass of $114\,\mathrm{M_{\earth}}$ and an
unusually large density. Calculations by the discovery-team give the
most likely core mass as $67\,\mathrm{M_{\earth}}$.

We will show that the large core of \hd\  cannot be explained by
migration. It has to form in-situ to allow such a large
subcritical core-mass.

\section{Modelling the equilibrium envelope structures}
\label{sec:modell-equil-envel}
Every protoplanet in our model consists of a solid core of constant
density ($\rho_\mathrm{core}$$=5500$$\,$kg\,$\mathrm{m}^{-3}$)\footnote{a core density of 10500 gives similar
 results} and an envelope of
hydrogen 
and helium with a helium mass fraction of 0.24. The composition of the envelope is assumed to
			be protosolar. Following the recommended procedure
			for the SCVH equations of state, that contain now high-Z
			elements, their effects are accounted for by a
			somewhat enhanced He-mass fraction. Heavy elements
			and their condensates are fully taken into account
			in the opacities to calculate radiative transfer
			efficiency in detail.

\subsection{Model equations and assumptions}
The hydrostatic equilibrium configuration of the envelopes is given by
the well known equations of stellar structure:
\begin{eqnarray}
  \label{eq:alleq}
  \frac{\mathrm{d}M}{\mathrm{d}r}&=&4\pi r^2 \rho(P,T),\\
  \frac{\mathrm{d}P}{\mathrm{d}r}&=&-\frac{G M}{r^2}\rho(P,T),\\
  \frac{\mathrm{d}T}{\mathrm{d}r}&=& \frac{T}{P} \frac{\mathrm{d}P}{\mathrm{d}r} \nabla(P,T),\\
  \frac{\mathrm{d}L}{\mathrm{d}r}&=&0,\\
  \frac{\mathrm{d}U}{\mathrm{d}r}&=&\frac{G M}{r^2},
\end{eqnarray}
This, together with the boundary conditions (see
section~\ref{sec:boundary-conditions}), determines the values for the
mass $M$, the pressure $P$, the temperature $T$, the luminosity $L$,
and the gravitational potential $U$ as a function of the radius
$r$. $G$ is the gravitational constant and $\nabla$ is calculated as:
\begin{equation}\label{eq:nabeff}
  \nabla = {\rm min}_{\rm smooth}(\nabla_\mathrm{rad}, \nabla_\mathrm{s}),
\end{equation}
i.e. the adiabatic temperature gradient $\nabla_\mathrm{s}$ or the
temperature gradient as caused by radiative energy transport in the
diffusion approximation, $\nabla_\mathrm{rad}$ -- whichever is smaller.
This corresponds to the use of zero entropy
			gradient convection and the application of the
			Schwarzschild-criterion, but is continuously differentiable across the
transition region. $\nabla_\mathrm{s}$ is directly given by the equation of
state, $\nabla_\mathrm{rad}$ is calculated as:
\[
  \label{eq:radiation_diffusion}
 \nabla_\mathrm{rad} =  \frac{3}{64\pi \sigma G}\, \frac{\kappa L P}{T^4 M}
\]
\citep[see][chapter 6 for a derivation of this formula and explanation
of all parameters]{mihalas}. $\kappa$ is the Rosseland-mean opacity and is
determined by using tabulated values (see sect. \ref{sec:opacity}).

\subsection{Boundary conditions}
\label{sec:boundary-conditions}
To solve the differential equation system we specify the following
boundary conditions:

\begin{itemize}
\item The core radius is defined using a fixed core density $\rho_\mathrm{c}$: 
\[  r_\mathrm{core}=\sqrt[3]{\frac{3}{4\pi}\frac{M}{\rho_\mathrm{c}}}.\]
\item The outer radius is given by the Hill radius: 
\begin{equation}
  r_\mathrm{Hill}= a \sqrt[3]{\frac{M}{3 M_\ast}},\label{eq:1}
\end{equation}
where $a$ is the planet's semi-major axis, $M$ its mass, and $M_\ast$ the
mass of the host star.
\item The luminosity $L$ is defined as the energy libration rate obtained for a constant
		planetesimal accretion rate and the dissipation of the planetesimal
		kinetic energy at the core surface:
\begin{eqnarray}  \label{eq:L}
  U(r_\mathrm{core})&=&-\int_{r_c}^{r_\mathrm{Hill}} \frac{G M_r}{r^2} dr,\quad
  \mathrm{and}\nonumber\\
  L&=& -(U-U_0) \dot{M},
\end{eqnarray}
where $U_0$ is arbitrary. $U_0$ is the gravitational potential at $r_\mathrm{Hill}$.
\end{itemize}

\subsection{Constituent relations}
\label{sec:const-relat}
To fully specify the differential equation system,
$\nabla_\textrm{s}$, $\rho{}$, and $\kappa$ need to be specified. We use the
following equation of state and opacity tables.
\subsubsection{Equation of state}
$\nabla_\textrm{s}(P,T)$, and $\rho{}(P,T)$ are interpolated from
\citet{1995ApJS...99..713Sv}. First hydrogen and helium are
interpolated independently, then the mixed quantity is determined. $\nabla_\textrm{s}$ is calculated via
spline derivatives from the mixed entropy including the mixing entropy
term, as suggested in
\citet{1995ApJS...99..713Sv}. The helium mass-fraction is $Y=0.24$.

\subsubsection{Opacity}\label{sec:opacity}
Rosseland-mean opacities $\kappa(\rho,T)$ are interpolated from 
a combined table: Opacities include Rosseland-mean dust opacities
from \citet[][$\lg T \leq 2.3$]{1985Icar...64..471P}, \citet{1994ApJ...437..879A} values
in the molecular range, and \citet{1990ADNDT..45..209W} Los Alamos
high temperature opacities.

\section{Utilizing all equilibria to determine the environmental
  properties for \hd}
In order to solve the system of differential equations, a range of
parameters must be provided, namely the:
\begin{enumerate}
\item core mass $M_\mathrm{core}$,\label{item:1}
\item pressure at the core $P_\mathrm{core}$,\label{item:2}
\item mass of the host star  $M_\ast$,\label{item:4}
\item semi-major-axis of the planet $a$, \label{item:5}
\item nebula temperature $T_\mathrm{neb}$, and the \label{item:6}
\item planetesimal accretion rate $\dot M$.\label{item:3}
\end{enumerate}
Parameters \ref{item:1} and \ref{item:2} are our independent
parameters, they are varied in a scale-free way, i.e. equidistant in
the logarithm.

\ref{item:4},\ref{item:5}, and \ref{item:6} are determined by the host
star and the position of the planet. The nebula temperature can be calculated
in thermal equilibrium with the star: 
\begin{equation}
  \label{eq:nebT}
  T_\mathrm{neb} = 280\cdot \left(\frac{a}{1 \mathrm{AU}}\right)^{-1/2} \left(\frac{L_\ast}{L_\odot}\right)^{1/4} \, \mathrm{K}
\end{equation}
\citep[see][]{1981PThPS..70...35H,1985prpl.conf.1100H}.

The only remaining free parameter is the planetesimal accretion rate
$\dot M$. In the case of \hd\ we choose an unusually large number of 
 $\dot M=10^{-2}\,M_{\earth}\, a^{-1}$. This value corresponds to
 particle-in-a-box planetesimal accretion for the density at the
 position of \hd\ in a 
 minimum mass solar nebula
 \citep{1981PThPS..70...35H,1985prpl.conf.1100H} with a gravitational
 enhancement factor $F_\mathrm{g}=21$.

\begin{figure*}
\centering
\includegraphics[width=\textwidth]{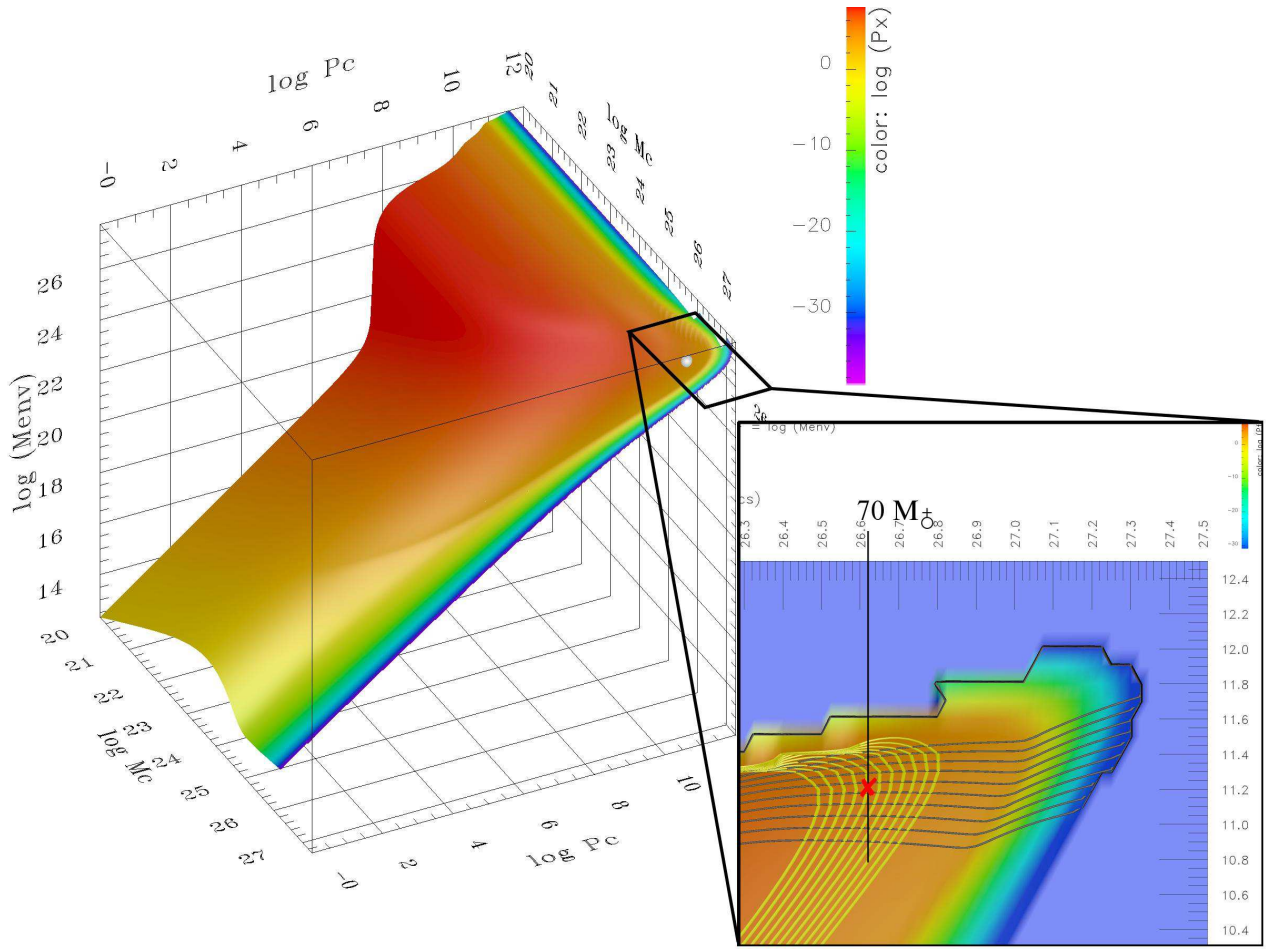}
\caption[Hydrostatic models for \hd.]{The figure shows the envelope mass ($\log M_\mathrm{env}$) as a function of core mass  and
pressure at the core surface (in the figure $\log \mathrm{Mc}$ and $\log \mathrm{Pc}$,
respectively). The logarithms are taken from the corresponding values
in SI units (kg / Pa). The results are given as a surface in three
dimensions and the surface colour is mapped from the outside pressure
($\log \mathrm{Px}$). The blow-up on the right-hand-side shows the
region of interest around a core mass of $70\,M_{\earth}$.\\
To determine the nebula conditions for a critical core mass of
$70\,M_{\earth}$, we have drawn lines of constant nebula pressure
(isobars, yellow, in the range $\lg P\, \mathrm{Pa}^{-1}=3..4$, step
0.1) and lines of constant envelope mass (gray, in the range $\lg
M_\mathrm{env}\, \mathrm{kg}^{-1}=26..27$, step 0.1). The critical core
mass for a given nebula pressure is given as the largest possible core
mass for a given isobar. So the isobar that is tangentially touched by
the line of $70\,M_{\earth}$ determines the nebula pressure for which
the critical core mass is $70\,M_{\earth}$. It is
$\lg P\, \mathrm{Pa}^{-1}=3.6$. \\
Using the gray lines we can immediately give the envelope mass
corresponding to this critical core mass. It is $47\,M_{\earth}$ giving
a total mass of roughly $117\,M_{\earth}$. This is almost exactly the
total mass of \hd.}
\label{fig:Hd_huelle}
\end{figure*}

For \hd\ the correct values are therefore:
\begin{itemize}
\item $M_\ast = 1.3\, M_\mathrm{\sun}$, $L_{\ast} = 2.72\,L_{\sun}$,
\item $a=0.042\,$AU,
\item $T_\mathrm{neb}=1754\,$K,
\item $\dot M=10^{-2}\,M_{\earth}\, a^{-1}$.
\end{itemize}

Using these values we have calculated all hydrostatic envelope
structures by solving the equation system described in
chapter~\ref{sec:modell-equil-envel} for a wide range of core masses
and pressures. Using this solution  manifold it is easy to determine
the correct environment that allows such a large critical core mass of
roughly $70\,M_{\earth}$ (see figure~\ref{fig:Hd_huelle}).

The results show that a hydrostatic equilibrium of gaseous envelope
and solid core is indeed possible at the specified position, if the
nebula pressure is $P=10^{3.6}\,\mathrm{Pa}=3980\,\mathrm{Pa}$. In
that case the envelope mass is  $\lg
M_\mathrm{env}\, \mathrm{kg}^{-1}=26.45$ or roughly $47\,M_{\earth}$
leading to a total of $117\,M_{\earth}$ for the
protoplanet.\footnote{Repeating these calculations with a core
  density of $\rho_\mathrm{core}=10500$$\,$kg\,$\mathrm{m}^{-3}$ and a
  critical core mass of $67\,M_{\earth}$ leads to
the same nebula pressure ($P=10^{3.6}\,\mathrm{Pa}$), a slightly
larger envelope mass ($\lg
M_\mathrm{env}\, \mathrm{kg}^{-1}=26.5$, or $53\,M_{\earth}$) and a
total mass of $120\,M_{\earth}$.} This is very close to the observed
mass of $114\,M_{\earth}$.

It should be noted, that this is the case only very close to the host
star. With increasing distance, the accretion rate $\dot M$ decreases,
and even
more importantly the hill radius increases for constant planet
mass. Both effects lead to a reduced critical core mass. Therefore,
such a large core is only allowed for very close-in planets like \hd.

\section{Fluid-dynamic formation of \hd}
\label{sec:fluid-dynam-form}
Knowing the environmental conditions, especially the values of the
nebula pressure $P_\mathrm{neb}$ and accretion rate $\dot M$ we tried
to reproduce the in-situ formation of \hd\ with a thorough fluid-dynamic
calculation. 

The algorithm for the dynamical calculations is described in
\citet{1990A&A...238...83W,1991Icar...91...39W,1991Icar...91...53W}.
For this paper we use 
a modified convection theory as in \citet{2003A&A...398.1081W}. 
These
calculations use a different equation of state \citep[not SCVH,
see][]{1989DissWuchterl}
and slightly different dust opacities, namely
			 for interstellar dust instead of the Pollack et al.
			 protosolar dust, see \cite{2003A&A...398.1081W}.

Because the dynamical calculation uses a particle-in-a-box accretion
scheme instead of a constant accretion rate, we use a gravitational
enhancement factor $F_\mathrm{g}=21$ (or a Safronov number $\theta=10$)
which leads to a peak accretion rate of $\dot M=10^{-2}\,M_{\earth}\,
a^{-1}$ as required.

\begin{figure}
\centering
\includegraphics[width=\columnwidth]{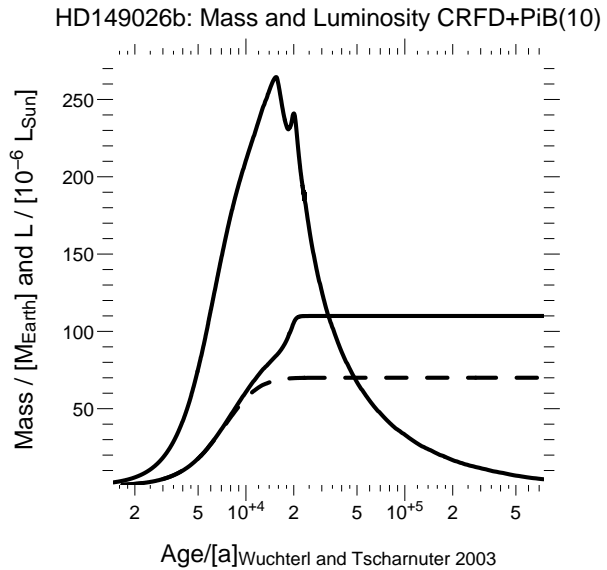}
\caption[Evolution of \hd]{Possible evolution of \hd. The figure shows
  the time-evolution of three quantities: the core mass (dashed), the
  total mass (solid), and the luminosity (solid). The age is defined
  as in \citet{2003A&A...398.1081W}, CRFD+PiB(10) stands for fully
  hydrodynamic simulation with Safronov number 10. The smooth
  luminosity curve is a strong hint for quasistatic evolution which is
  confirmed by looking at the data. The dynamic terms are negligible.}
\label{fig:hd_wuchterl}
\end{figure}

In spite of the slight differences regarding the constituent relations, the
dynamic calculations confirm the hydrostatic model. The evolution of
\hd\ is plotted in figure~\ref{fig:hd_wuchterl}.
The dynamic calculation confirms quasistatic evolution for the entire
formation process of \hd, no instabilities occur. This formation
scenario explains the high core mass and shows no dynamic
accretion phase -- the planet grows hydrostatically all the way to its
final mass.

\section{Solving the feeding-zone problem}
\label{sec:solving-feeding-zone}
So far we have shown how \hd\ could have formed provided that there is
enough material available to form the planet at its current
position. The lack of building material is usually considered the
strongest argument against in-situ formation of close-in planets.

It is true that in a classical feeding zone, i.e. 3--4 hill radii
(equation~\eqref{eq:1}) on 
both sides of the orbit of the planet, there is not enough material
available in a gravitationally stable disk. This problem can be
circumvented by assuming a continuous flow of material onto the star
as is the case for accretion disks. According to
\citet{1998ApJ...495..385H} typical T~Tauri disks with an age of 1\,Ma
have relatively low accretion rates, they say:
\begin{quotation}
"The median accretion rate for T Tauri stars of age $\sim1$\,Ma is
$10^{-8}\,M_{\sun}\, \textrm{a}^{-1}$; the intrinsic scatter at a 
given age may be as large as 1 order of magnitude."
  \end{quotation}

For now, we will assume a very fast formation of the planet of $10^5\,$a and
calculate the amount of material that passes the orbit of the
hypothetical in-situ planet. Using Hartmann's estimate, the material
passing a close-in planet during its formation is therfore $M=10^5\times
10^{-8}\,M_{\sun}=333\,M_{\earth}$. For the envelope this is quite
enough mass, but what about the heavy element core? For solar
composition gas we have a mass-fraction of condensible material of $\approx1/56$
outside of the ice-line and $\approx1/240$ inside
\citep{1985prpl.conf.1100H}. Typical disk-lifetimes are  $\sim1$\,Ma. We
can now calculate the amount of condesible material that passes the
protoplanet's orbit during the lifetime of the disk:
\begin{equation}
  \label{eq:19}
  M_\mathrm{cond} = 10^6\times 10^{-8}M_{\sun} / 240 \approx 14 \,M_{\earth}.
\end{equation}
Keeping in mind that \hd\ is an extreme case, it is quite appropriate
to use the upper limit given by Hartman: a disk accretion rate of
$10^{-7}\,M_{\sun}\, \textrm{a}^{-1}$. This provides $140\,M_{\earth}$
of condensible material at the orbit of the planet. 

In the case of \hd\ we can go even further. \hdA's metalicity is
given as [Fe/H]$=$0.36 \cite[from][]{2005ApJ...633..465S}. Assuming
that the other heavy elements are similarly enriched, this is an
overabundance in heavy elements of $10^{0.36}=2.3$. In total we end
up with $\sim300\,M_{\earth}$ in heavy elements passing the planet's
orbit -- This is enough solid material for \hd.

There is one problem remaining, concerning the condensibles: the high
temperature of the nebula. At a temperature of 1754\,K and a pressure
of 4000\,Pa will there be any condensed specimen left? According to
\citet{1996A&A...312..624D} the mass fraction of 
silicon for this $P$-$T$-regime is 0.99-0.999. So the silicates are
still available. For other specimen, especially carbon, this is
usually not the
case. To answer the question of condensible mass-fraction precisely,
the exact environmental conditions, such as pressure,
temperature, and chemical composition at the time of \hd's
formation need to be taken into account. Without this information, we
can only speculate. As it is very
likely that cm-sized and larger grains spiral inward to the star, a
fraction of the particles will never reach equilibrium before meeting
the protoplanet. Once inside the planet, the high pressure environment
prevents evaporation. Therefore we assume that a fraction of the
non-silicates can also be accreted onto the core of the protoplanet
providing just enough material to form \hd.



%

\section{Notes on Migration}
\label{sec:notes-migration}

The reader might have pondered the lack of migration in the above
discussion -- this is partly intended. We wanted to show that while different
types of migration can take place for different embryo masses and
nebula proberties, they are not strictly necessary for the formation
of close-in planets.

In the case of \hd\ we can go even further: Our calculations show that
to obtain such a large core, at least the last phase of core accretion
must have occured in close proximity to its present location,
i.e. without migration. The 'last
phase' in this case refers to the time when the core mass grows beyond
$\sim30\,M_{\earth}$. Before that time, migration could have occured --
this is irrelevant for the formation scenario presented here. Once the
planet nears its final core mass, the nebula pressure is already very
low. Therefore the planet's orbit should be stable against type-I
migration.

This scenario has some interesting consequences. The
conventional accretion model for planet formation at larger
separations predicts a typical core size. Two giant planets forming in
the same nebula should be of roughly similar mass, i.e. giant planets
orbiting the same star can be expected to have similar core
masses. This is not true for the in-situ formation scenario presented
in this paper. In the case of \hd\ we could show that there is no
dynamically triggered gas-accretion phase that sets in beyond a
critical core mass. If this is the case for all close-in planets,
core masses will  only be determined by the amount of
available material. Therefore we expect a wide distribution of core
masses for close-in planets in the case of in-situ formation. This
property could be used to distinguish between in-situ formation and
large-scale migration.

\section{Conclusions}
\label{sec:conclusions}
We have demonstrated a plausible formation scenario for \hd. In-situ
formation permits the growth of the planet's very large core up to its present size.

Provided that enough material is available, even very low densities
(e.g. minimum mass solar nebula values), produce very high planetesimal
accretion rates. A safronov number of 10 is enough to produce
accretion rates of $\dot M=10^{-2}\,M_{\earth}\, a^{-1}$ building a
$70\,M_{\earth}$ core in less than 10000\,a. By analysis of all
possible hydrostatic envelopes around such a large core we could
determine the correct nebula pressure and demonstrate a fluid-dynamic
model for the formation of \hd. 

In chapter~\ref{sec:solving-feeding-zone} we analysed the mass flow in
a typical T~Tauri star accretion disk at an age of 1~Ma. We could
show that for such a case, enough material is transported across the
orbit of a close-in planet like \hd\ to provide sufficient material
even when considering grain-evaporation caused by the high temperatures 
 this close to the star.

The planet \hd\ could indeed have formed in-situ, i.e. at its present
position. In that case its entire evolution would have been
quasi-static -- no dynamic accretion takes place. Forming \hd\ at
larger distances in this manner -- e.g. outside the ice-line -- is not possible as the
critical core mass decreases dramatically with increasing distance.

Our model predicts a wide range of core masses for close-in planets
while migrating planets should all have similar 'typical' critical
cores. This property could be used to distinguish between the two
formation scenarios.

\section*{Acknowledgements}
This research was supported in part by DLR
project number 
50-OW-0501.

\bibliographystyle{mn}
\bibliography{lit/Literatur}

\label{lastpage}
\end{document}